\documentstyle[itzidef]{mn}

\input epsf

\title[Adaptive Optics observations of LBQS 0108+0028
]
{Adaptive Optics observations of LBQS 0108+0028:
$K$-band detection of the host galaxy of a radio-quiet QSO at $z\approx2$}

\author[I. Aretxaga, D. Le Mignant, 
J. Melnick, Terlevich, Boyle B.J.]
       {Itziar Aretxaga$^1$, D. Le Mignant$^2$, 
J. Melnick$^2$,  R. J. Terlevich$^3$, B.J. Boyle$^4$ \\
$^1$ Max-Planck-Institut f\"ur Astrophysik, Karl Schwarzschildstr. 1,
Postfach 1523, 85740 Garching, Germany\\
$^2$ European Southern Observatory, Alonso de Cordova 3107,
Vitacura, Casilla 19001, Santiago 19, Chile.\\
$^3$ Royal Greenwich Observatory, Madingley Road, Cambridge CB3 0EZ, U.K.\\
$^4$ Anglo-Australian Observatory, PO Box 296, Epping, NSW 2121 Australia}
\begin{document}

\maketitle

\begin{abstract}
We report the {\em first} unambiguous detection of the host galaxy of
a normal radio-quiet QSO at high-redshift in $K$-band. The luminosity
of the host
comprises about 35\%\ of the total $K$-band luminosity.
Assuming the average colour of QSOs at $z\approx2$, the host would be
about 5 to 6~mag brighter than an unevolved $L_*$ galaxy placed at 
$z\approx2$, and 3 to 4~mag brighter than a passively evolved 
$L_*$ galaxy at the same redshift.
The luminosity of the host galaxy of the QSO  would thus overlap with the
highest found in radio-loud QSOs and radio-galaxies at the same redshift.

\end{abstract}

\begin{keywords}
galaxies: active -- galaxies: photometry --
galaxies: quasars: general  
\end{keywords}

%%%%%%%%%%%%%%%%%
%%%% Introduction
%%%%%%%%%%%%%%%%%

 \begin{table*}
 \begin{minipage}{140mm}
 \begin{center}
  \caption{Summary of observations}
  \begin{tabular}{lccccccl}
   Object     & $\alpha$ (J2000)  &  $\delta$ (J2000) & $z$ & $V$(obj)
& $\theta$(obj - $\star$) & $V(\star)$ & $t$ \\
              &   & &   & (mag)  & (arcsec) & (mag) & (s)   \\
LBQS 0108$+$0028 & {1} {10} {38.1} & {0} {44} {54} & 2.005 & 18.3
 & $21.0$ & $12.0$ & 10800 \\
star-A        & {1} {4} {1.24} & {0} {55} {6.7}  & -- & 13.6 & 20.6 & 11.3 & 
5400 \\
 \end{tabular}
\end{center}
\end{minipage}
\end{table*}

\section{Introduction}

Recent evidence that the cosmological evolution of the 
density of star formation in the Universe  (Madau et al. 1996)
follows closely the QSO density evolution (Boyle \& Terlevich 
1998) emphasizes the need to study the kinds of galaxies that host
Active Galactic Nuclei in order to understand the link between 
star-formation and nuclear activity, and potentially the role of 
nuclear activity in galaxy formation. 

At the peak value of 
QSO density ($z\approx 2$ to 3),
the few QSO host-galaxies detected so far present rest-frame 
UV fluxes that reach
up to 20\%\ of the total QSO luminosity,
indicating star-formation rates about 200 \Msun/yr and above for both
radio-loud (Lehnert et al. 1992) and radio-quiet samples (Aretxaga, Boyle
\& Terlevich 1995, Hutchings 1995). These values are almost an order 
of magnitude above those of field galaxies at similar redshifts
selected through Lyman Break techniques (Steidel et al. 1996,
Lowenthal et al. 1997). The properties of these QSO hosts are not 
unprecedented, since they
follow very closely the luminosity--size relation of nearby
star forming galaxies, overlapping with its high-luminosity end 
(Aretxaga, Terlevich \& Boyle 1998).
However, the UV fluxes only carry information
about the high-mass end of the stellar populations in the galaxies,
and say little about the bulk of the stellar mass which is better 
characterized by optical to NIR observations. 
%A clear link between
%QSO hosts and other types of galaxies is therefore to be investigated.

Although a few hosts of extreme radio-loud QSOs at $z\approx 2$
have been detected in NIR
bands (Lehnert et al. 1992, Carballo et al. 1998),
attempts to image the hosts of normal radio-quiet
QSOs at the same redshifts
have been unsuccessful to date 
(Lowenthal et al. 1995, Aretxaga et al. 1998). 
Imaging radio-quiet systems, which constitute 
more than 95\%\ of all QSOs, is important in order to 
characterize the bulk of the population.
The observed optical sizes of
FWHM$\approx 1$ arcsec (Aretxaga et al. 1995), clearly demand a technique 
which offers the highest available angular resolution.

In this paper we focus our attention on the detection of the host
of a normal radio-quiet $z\approx 2$ QSO with the Adaptive Optics System in 
operation at the ESO 3.6m telescope in La Silla. Preliminary results
on similar programs to image the host-galaxies 
of QSOs at $z\approx 0.5 \hbox{ \ and \ } 1.7$ 
using Adaptive Optics have been presented in a recent
conference devoted to quasar hosts (Bremer et al. 1997, Hutchings 1997).

%%%%%%%%%%%%%%
%%%% Section 2
%%%%%%%%%%%%%%

\ifoldfss
  \section{Data set: Acquisition and Reduction}
\else
  \section[]{Data set: Acquisition and Reduction}
\fi

We selected  LBQS~0108+0028 at \RA{J2000}{1}{10}{38.1} \DEC{J2000}{0}{44}{54},
a $V=18.3$~mag QSO at $z=2.005$ which was 
discovered in the Large Bright Quasar Survey (Hewett, Foltz \& Chaffee 1995),
because it belongs to a narrow redshift-luminosity 
band ($1.8 \lsim z \lsim 2.2$, $M_B \lsim -28$ mag for \Ho50\ and \qo0p5)
and it lies close in projection $\theta = 21$~arcsec to 
a bright star of magnitude $V=12.0$~mag. The first selection criterion
%was imposed in order to overlap with QSOs already studied by some of us.
was imposed in order to explore the luminosity band that 
is predicted to contain the most luminous hosts by quasar formation theories 
(Terlevich \& Boyle 1993, Haenelt \& Rees 1993), and it
has indeed provided a high detection rate of extended fuzz (Aretxaga
et al. 1995).
The second condition was 
imposed in order to be able to correct the atmospheric turbulence 
with Adaptive Optics, using a nearby
bright reference star on-axis,
since QSOs at these redshift are too faint to allow for direct 
corrections on themselves.
There is no radio-detection of 
this QSO. 

The observations were carried out in $K'$-band 
on six hours spread through the nights
of 1995 October 10th, 11th and 12th at the ESO 3.6m
telescope in La Silla, with {\sc Come-On+}
%the Adaptive Optics Near Infrared System 
(Rigaut F. et al. 1991, Rousset et al. 1994). 
%that was later upgraded to {\sc adonis}.
We used Sharp II, the 256x256 Nicmos III array, in the general purpose
50 marcsec/pixel resolution mode, which gives a 12.8 x 12.8~arcsec$^2$
field of view.

In order to accurately measure a realistic 
Point Spread Function (PSF) we observed every night a comparison double system 
that mimics the 
brightness of the reference star and its relative distance to the QSO.
The comparison system comprises a  star of magnitude $V=13.6$ at 
\RA{J2000}{1}{4}{1.24} \DEC{J2000}{0}{55}{00.6}, 
which we denote A,  separated by 
$\theta = 20.6$~arcsec
from a star B of magnitude $V = 11.3$~mag at
\RA{J2000}{1}{4}{1.24} \DEC{J2000}{0}{55}{6.7}.
Star B was used as reference star to correct for atmospheric turbulence
and star A was imaged to serve as a PSF calibrator star 
in the analysis of the QSO profile.

Coordinates, redshift, $V$-band brightnesses, distances from reference stars,
brightnesses of reference stars, and total integration times are
summarized in Table~1.

  The observations were carried out in the following sequence:
PSF-star (5 x 30 sec) in position 1 --  
PSF-star (5 x 30 sec) in position 2 --
QSO (10 x 60 sec) in position 1 -- QSO (10 x 60 sec) in position 2 --
PSF-star (5 x 30 sec)  in position 1 ... in a repeating cycle
totaling 3 hr of integration for 
the QSO and 45 min for the PSF-star.
Different frames were offset by approximately $6$~arcsec (distance between 
positions 1 and 2) from one another 
in order to estimate the sky level from contiguous frames.

The seeing, as recorded by the
Differential Image Motion Measurement, 
was $0.8$~arcsec during the first night, 
being very stable ($\pm 0.03$~arcsec), but variable the second 
and third nights ($0.8$ to $1.8$~arcsec). 

%\ojo\ {\it David, please check and correct:}

  The data were reduced with the image processing package 
{\bf eclipse}\footnote{{\bf eclipse} is an image processing engine developed 
at ESO for astronomical data reduction purposes in general, and 
adaptive optics data reduction in particular (also, see the web site 
{\tt http://www.eso.org/eclipse/})} (Devillard N. 1997).  
%working in  IRAF\footnote{IRAF is distributed by
%the National Optical Astronomy Observatories, which are operated by
%the Association of Universities for Research in Astronomy, Inc., under
%cooperative agreement with the National Science Fundation}. 
The data was
first sky subtracted, using an average of contiguous frames with 
misplaced sources,
and then flat-fielded with a gain-corrected sky flat frame. 
Bad pixels were identified in the gain map and substituted by linearly 
interpolated values from nearby pixels. The shift-and-add routines of 
{\bf eclipse} were then
used to register individual frames and coadd them
in imaging stacks of 10~min for QSO frames and 5 min for PSF 
calibrator star frames.

The final FWHM of the coadded stacks of the PSF calibrator star 
were $0.3$ to $0.4$ 
arcsec during the first night; $0.3$ to $0.9$ arcsec during the second night;
and $0.7$ to $1.0$ arcsec during the third night. The Strehl ratios attained 
(8 to 12\% for Oct.10th, but below 5\% for Oct.11th and 12th)
 were only acceptable during the first night. 
The second and third night image quality was poor, 
partly because fast sky variations and 
partly because fast bad seeing, 
which provoked a reduction 
of the isoplanatic patch to distances much smaller than those of our object to
reference star systems.

The images were not flux-calibrated since we didn't acquire
calibration stars and there isn't any $K$-band measurement of this
QSO in the literature.

%%%%%%%%%%%%%%
%%%% Section 3
%%%%%%%%%%%%%%

\ifoldfss
  \section{Profile analysis}
\else
  \section[]{Profile analysis}
\fi

\begin{figure*}
    \cidfig{5in}{26}{144}{580}{456}{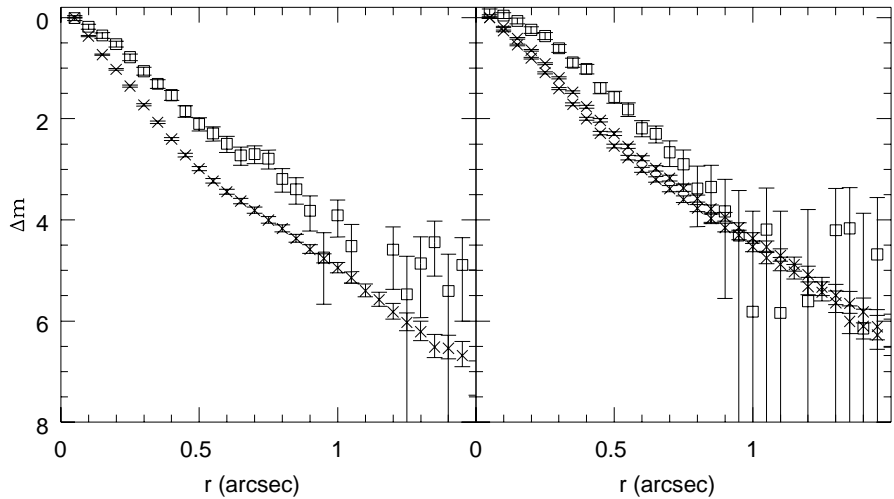}
    \caption{Radial profiles of two QSO coadded images of 20~min
    exposure (squares),
compared to the stellar profiles (stars) acquired before and after the
QSO during the first night. Fluxes are normalized to their centroid values}
\end{figure*}

For each 10~min QSO images and 5~min PSF-star images, we derived an 
azimuthally averaged radial profile using the STSDAS package in 
IRAF\footnote{IRAF is distributed by
the National Optical Astronomy Observatories, which are operated by
the Association of Universities for Research in Astronomy, Inc., under
cooperative agreement with the National Science Fundation}. 
We checked for variations across the chip (position~1 versus position~2)
using the PSF profiles obtained during the first night, when 
the seeing
was stable. We detected no variations within the error bars,
and therefore coadded contiguous stacks of QSO and PSF-star frames, into 
20~min and 10~min exposure images.

A comparison of the QSO profile and the PSF profile, normalized in order
to reproduce the luminosity of the QSO at its centroid, are however
different within the inner arcsec. Figure~1 represents
two sets of QSO profiles versus the PSF-star profiles obtained before
and after each QSO observation. The recorded profiles were 
very stable throughout the night.

During the second and third nights the QSO profiles are typically 
indistinguishable from those of the PSF-star and, when different,
they are enclosed by the varying seeing profiles. Figure~2 shows the
comparison between QSO and PSF-star for one of these cases. As stated
in the previous section, during these nights the correction attained
by {\sc Come-On+} was poor due to the poor seeing conditions, and there was 
little improvement in the spatial resolution of the images. The resulting
stellar FWHM $\approx 1.0$ arcsec 
implies that we shouldn't expect to have resolved 
the extended structures at $r< 1.0$ arcsec that we 
detected during the first night,
when the seeing was good and stable.

%\begin{figure}
%    \cidfig{2.5in}{24}{145}{300}{430}{QSOvsQSO.ps}
%    \caption{Radial profiles of two QSO images taken during the
%first night, normalized to their centroid values.}
%\end{figure}

\begin{figure*}
    \cidfig{2.8in}{24}{145}{300}{430}{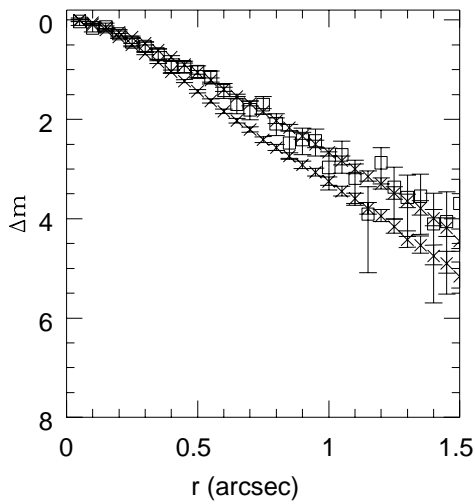}
    \caption{As Figure~1, for the second night.}
\end{figure*}

\section{Results}
The QSO profile is more extended than that of the PSF-star, 
with $K'$ excesses of about 35\%\ the total luminosity of the QSO.
This is the first clear detection of the host of a distant
radio-quiet QSO
in the NIR.
Previous attempts to detect them 
with direct imaging (Lowenthal et al. 1995, Aretxaga
et al. 1998) resulted in non-detections at $z\approx 2$,
with the exception of a marginal detection of a
host at $r\approx4$~arcsec (Aretxaga et al. 1998). The
upper limits for the luminosities set by these studies were
an order of magnitude above our detection.
However, it is clear that this host would have gone undetected
by these studies, since all the signal is localized in the inner 1~arcsec
of the QSO profile while, typically, the previous studies were carried out with
 seeing values of FWHM between 0.9 and 1.3~arcsec (see also Fig.2).

If we assume that this QSO has the 
average colour of QSOs at $z\approx2$ $V-K \approx 2.2\pm 0.6$~mag 
(Hewett P.C., priv. communication), where the error bar indicates the
total amplitude of colours, 
then the $K$-band magnitude of the
extension would be about $17.2\pm 0.6$~mag, overlapping with the 
$K$-band apparent 
magnitudes of radio-loud QSOs (Lehnert et al. 1992, Carballo et al. 1998)
and radio-galaxies (Lilly 1989) at the same redshift.
Our $K$-band 
detection of the QSO host in this analysis
demonstrates that there is at least one good example of a radio-quiet
QSO host with an extremely luminous host galaxy at observed NIR wavelengths.

%The four radio-loud QSOs at $z\approx 2$ imaged
%by Lehnert et al. (1992), whose radio-emission is 
%typically extended on large scales ($\gsim10''$), are somewhat more
%luminous 
%in rest-frame UV than those of the
%radio-quiet  and 
%core-dominated radio-loud QSOs imaged 
%by Aretxaga et al. (1995, 1998) and Hutchings
%(1995). The only $z\approx 2$
%radio-loud QSO imaged by Aretxaga et al. (1995, 1998) has a core-dominated
%radio-emission, while the only $z\approx 2$ un-lensed radio-loud QSO
%imaged by Hutchings (1995) has point like structure at radio wavelengths
%(Douglas et al. 1996).
%detection points
%towards all kinds of luminous QSO hosts at $z=2$ having similar
%properties in 
%The enhancement in the rest-frame UV flux of lobe-dominated radio-loud
%QSOs could be attributed to the same phenomenon encountered in 
%radio-galaxies, where scattering or star-formation has been claimed
%to produce the observed alignment of the blue light with the radio-axis, 
%leaving the NIR bands
%almost unaffected (Rigler et al. 1992, Best, Longair \& R\"ottgering 1997).

Since at $z\approx2$ the observed $K$-band corresponds approximately to
rest-frame $R$-band, we can make an easy comparison of our host galaxy 
with local galaxies observed at optical wavelengths. A local 
$L_\star$ elliptical has a luminosity $M_R^*\approx -22.8$~mag, as
derived from the local luminosity function of field galaxies (Efstathiou,
Ellis \& Peterson
1988). Thus, the host of  LBQS 0108+0028
is likely to be about 
5 to 6~mag brighter than an unevolved $L_*$ galaxy placed at $z\approx2$. 
Taking into account the evolution that the stellar populations must
have experienced
between $z=2$ and $z=0$, 
an $L_*$ galaxy at $z\approx2$ would be about 2~mag 
brighter in $R$-band than
nowadays if it had been passively evolving since formation
(Charlot \&
Bruzual 1991). 
The host of  LBQS 0108+0028
would thus be 3 to 4~mag brighter than a passively evolved $L_*$ 
elliptical galaxy placed at $z\approx2$. 
Even higher luminosities should be considered if the light we
are missing near the center of the host is also taken into account.

The host of  LBQS 0108+0028 would also be at least 5~mag 
brighter than the average
radio-quiet host galaxies of nearby 
QSOs: \hbox{$<M_V>$} $\approx -21.6 \hbox{\ to \ } -22.6 $~mag for QSOs at 
$<z>\approx 0.2$
(e.g. Smith et al. 1986, Hutchings, Janson \& Neff 1989, Bahcall et al. 1997) 
which with a 
$V-R \approx 0.7$~mag colour for an Sb to E galaxy (Fukigita, Shimasaku \& 
Ichiwava  1995)
gives rest-frame luminosities of $M_R\approx -22.2 \hbox{\ to \ } -23.2$~mag.
Nearby IRAS selected QSOs can reside
in very luminous galaxies of up to $L\sim 6 L_\star$ (Boyce et al. 1996).

As already noted by Lehnert and co-workers (1992)
for their radio-loud sample,
the density of luminous QSOs ($M_B \lsim -28$ mag) at $z\approx 2$
like the one explored in this study
is about 10 Gpc$^{-3}$ (Boyle et al. 1991), and their hosts 
can be well accommodated at $z=0$ by the tail of the 
luminosity function of field galaxies,
an idea which has also been proposed by Terlevich and Boyle (1993) in their
comparative study of the luminosity functions of QSOs and elliptical galaxies.
  Clearly, a bigger sample of hosts of radio-quiet QSOs should be 
detected in $K$-band
before establishing an evolutionary link between these populations
of galaxies.

\section*{Acknowledgments}
We thank S.D.M. White 
for providing useful
comments on an earlier draft of this paper. 
This work was started thanks to the visiting program of ESO.
IA and RJT, who benefited from it, acknowledge the kind hospitality of the 
astronomers and other 
members of staff during their visit at ESO, Santiago (Chile).
This work has also been supported by the `Formation and Evolution of
Galaxies' network set up by the European Commission under contract 
ERB FMRX-CT96-086 of its TMR programme. 
%This research has made use of the NASA/IPAC Extragalactic Database (NED)
%which is operated by the Jet Propulsion Laboratory, California Institute of
%Technology, under contract with the National Aeronautics and Space
%Administration. 

\end{document}